# Classification of Heart Disease Using K- Nearest Neighbor and Genetic Algorithm

M.Akhil jabbar* B.L Deekshatulu[a] Priti Chandra [b]

*Associate Professor, Aurora's Engineering college, Bhongir, A.P, INDIA*
[a]*Distinguished Fellow IDRBT, INDIA*
[b]*senior scientist, ASL, DRDO, INDIA*

**Abstract**

Data mining techniques have been widely used to mine knowledgeable information from medical data bases. In data mining classification is a supervised learning that can be used to design models describing important data classes, where class attribute is involved in the construction of the classifier. Nearest neighbor (KNN) is very simple, most popular, highly efficient and effective algorithm for pattern recognition. KNN is a straight forward classifier, where samples are classified based on the class of their nearest neighbor. Medical data bases are high volume in nature. If the data set contains redundant and irrelevant attributes, classification may produce less accurate result. Heart disease is the leading cause of death in INDIA. In Andhra Pradesh heart disease was the leading cause of mortality accounting for 32% of all deaths, a rate as high as Canada (35%) and USA. Hence there is a need to define a decision support system that helps clinicians decide to take precautionary steps. In this paper we propose a new algorithm which combines KNN with genetic algorithm for effective classification. Genetic algorithms perform global search in complex large and multimodal landscapes and provide optimal solution. Experimental results shows that our algorithm enhance the accuracy in diagnosis of heart disease.

* Corresponding author. Tel.: +9912648686
  E-mail address: jabbar.meerja@gmail.com

## 1. Introduction

Data mining is the process of automatically extracting knowledgeable information from huge amounts of data. It has become increasingly important as real life data enormously increasing [1].Data mining is an integral part of KDD, which consists of series of transformation steps from preprocessing of data to post processing of data mining results. The basic functionality of data mining involves classification, association and clustering. Classification is a pervasive problem that encompasses many diverse applications. To improve medical decision making data mining techniques have been applied to variety of medical domains. Many health care organizations are facing a major challenge is the provision of quality services like diagnosing patients correctly and administering treatment at reasonable costs. Data mining techniques answer several important and critical questions related to health care.

Nearest neighbor is one of the most popular classification technique introduced by Hodges and fix [2].Without any additional data, classification rules are generated by the training samples themselves.

Evolutionary algorithms are used for the problem that can't be solved efficiently with traditional algorithms. Genetic algorithms (GA) are computing methodologies constructed with the process of evolution [3]. GA Have played a vital role in many engineering applications. Heart disease is the leading cause of death in developed countries and is one of the main contributors to disease burden in developing countries like INDIA. Several studies in Andhra Pradesh reveal that heart disease was the leading cause of mortality accounting 32% deaths, as high as Canada and USA.Hence there is a need to design and develop a clinical decision support for classification of heart disease. In this paper we propose a classification algorithm which combines KNN and genetic algorithm, to predict heart disease of a patient for Andhra Pradesh population.

This paper is structured as follows: section 2 we review the concepts of KNN, Genetic algorithm and heart disease. Section 3 explains our proposed classifier. Results are discussed in section 4 and we conclude our remarks in section 5.

## 2. Basic concepts

In this section we review the concepts like KNN, Genetic algorithm and heart disease.

### *2.1. K nearest neighbor classifier*

K nearest neighbor(KNN) is a simple algorithm, which stores all cases and classify new cases based on similarity measure.KNN algorithm also called as 1) case based reasoning 2) k nearest neighbor 3)example based reasoning 4) instance based learning 5) memory based reasoning 6) lazy learning [4].KNN algorithms have been used since 1970 in many applications like statistical estimation and pattern recognition etc.KNN is a non parametric classification method which is broadly classified into two types1) structure less NN techniques 2) structure based NN techniques. In structure less NN techniques whole data is classified into training and test sample data. From training point to sample point distance is evaluated, and the point with lowest distance is called nearest neighbor. Structure based NN techniques are based on structures of data like orthogonal structure tree (OST), ball tree, k-d tree, axis tree, nearest future line and central line [5].Nearest neighbor classification is used mainly when all the attributes are continuos.Simple K nearest neighbor algorithm is shown in figure 1

> **Steps 1)** find the K training instances which are closest to unknown instance
> **Step2)** pick the most commonly occurring classification for these K instances

Fig 1.K nearest neighbor algorithm

There are various ways of measuring the similarity between two instances with n attribute values. Every measure has the following three requirements. Let dist (A, B) be the distance between two points A,B then
1) dist(A,B)≥0 and dist(A,B)=0 iff A=B
2) dist(A,B)= dist(B,A)
3) dist(A,C)≤ dist(A,B)+ dist(B,C)

Property 3 is called as "Triangle in equality". It states that the shortest distance between any two points is a straight line. Most common distance measures used is Euclidean distance .For continuous variables Z score standardization and min max normalization are used [6].

KNN is used in many applications such as 1) classification and interpretation 2) problem solving 3) function learning and teaching and training.KNN suffers from the following drawbacks 1) low efficiency 2) dependency on the selection of good values for k.
Further research is required to improve the accuracy of KNN with good values of K.

*2.2. Genetic algorithm*

Evolutionary computing started by lifting ideas from biological theory into computer science. Genetic algorithms are most popular technique in evolutionary computing. Evolutionary algorithms are used in problems for optimization. To solve problems, evolutionary algorithms require a data structure to represent and evaluate solution from old solutions. Fig 2 shows flowchart of an evolutionary algorithm [7].

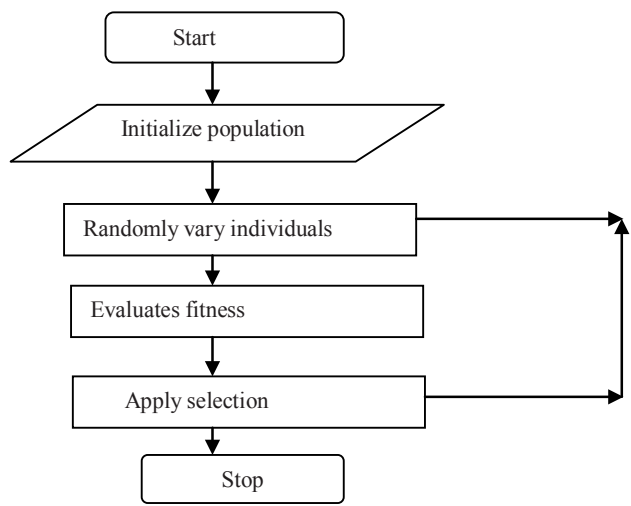

Fig 2. Flowchart of an evolutionary algorithm

Evolutionary algorithms are used in many applications such as
1) Machine intelligence
2) Travelling sales person problem
3) Expert system
4) Medicine
5) Engineering application
6) Wired and wireless communication systems

Genetic algorithms (GA) was invented by john Holland in 1975.Genetic algorithms are useful for search and optimization problems.GA uses genetics as it's model as problem solving. Each solution in genetic algorithm is represented through chromosomes. Chromosomes are made up of genes, which are individual elements (alleles) that represent the problem. The collection of all chromosomes is called population. Generally there are three popular operators are use in GA.

1) **Selection**:

This operator is used in selecting individuals for reproduction. Various selection methods are
i) Roulette wheel selection
ii) Random selection
iii) Rank selection
iv) Tournament selection
v) Boltzmann selection

2) **Crossover**: This is the process of taking two parent chromosomes and producing a child from them. This operator is applied to create better string

Various types of cross over operators are
i) Single point crossover
ii) Two point cross over
iii) N point crossover
iv) Uniform crossover
v) Three parent cross over
vi) Cross over with reduced surrogate
vii) Shuffle crossover
viii) Precedence preservative crossover
ix) Ordered crossover
x) Partially matched crossover

3) **Mutation**: This operator is used to alter the new solutions in the search for better solution. Mutation prevents the GA to be trapped in a local minimum [8].
4) **Fitness function**: fitness function in GA is the value of an objective function for its phenotype. The chromosome has to be first decoded, for calculating the fitness function.

Figure 3 shows process of genetic algorithm

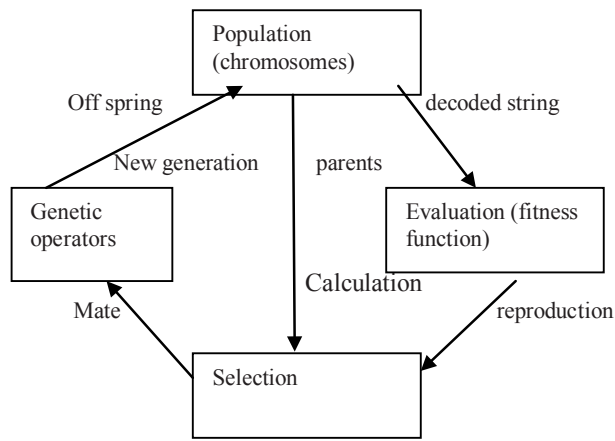

**Fig. 3 working of genetic algorithm**.

There are many advantages with genetic algorithm. Few of them are listed below
1) Solution space is wider
2) Easy to discover global optimum
3) Only uses function evaluations
4) Handles noisy functions well

Some limitations of GA are
1) Identifying suitable fitness function is difficult
2) GA require more number of fitness evaluations
3) These are not good to identify local optima
4) No straight forward configuration.

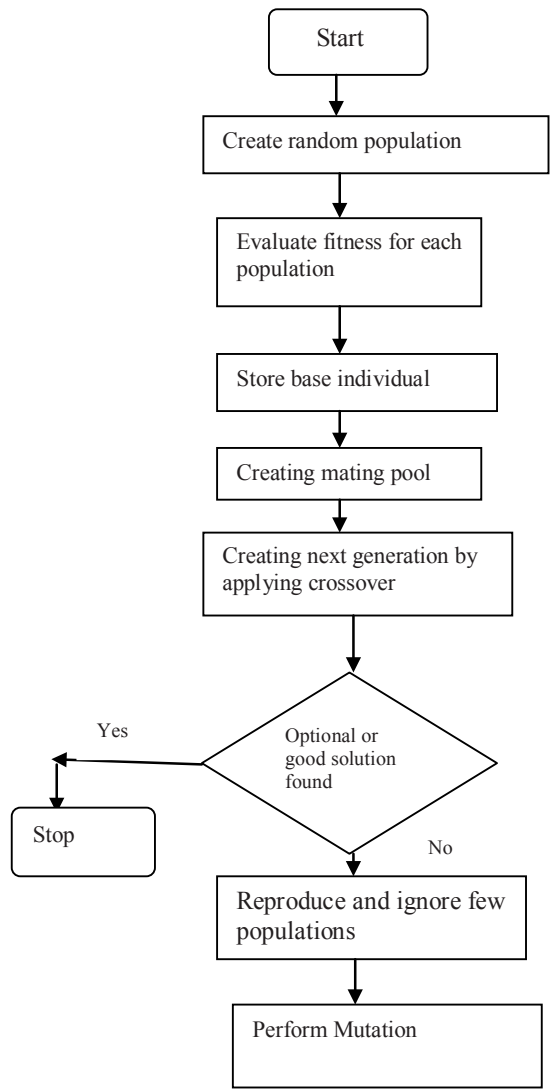

**Fig 4 flowchart of genetic algorithm**

*2.3. Heart Disease*

Heart disease occurs when the arteries which normally provide oxygen and blood to the heart blocked completely or narrowed. Heart problems acquired at birth or later in life. According to survey conducted by register general of India, heart disease is a major cause of death in India and Andhra Pradesh [9].

Various types of heart diseases are
1) Coronary heart disease
2) Cardiomyopathy
3) Cardiovascular disease
4) Ischaemic heart disease
5) Heart failure
6) Hypertensive heart disease
7) Inflammatory heart disease
8) Valvular heart disease

Common risk factors of heart disease include
1) High blood pressure
2) Abnormal blood lipids
3) Use of tobacco
4) Obesity
5) Physical inactivity
6) Diabetes
7) Age
8) Gender
9) Family history[10]

As per the survey conducted by WHO, out of 10 deaths in India, eight are caused by cardio vascular diseases and diabetes. Heart disease mortality in Andhra Pradesh is recorded as 30% [11]. Preventive strategies to reduce risk factors are essential and to reduce the alarmingly increasing burden of heart disease in our population.

**3. Proposed method**

Our proposed approach combines KNN and genetic algorithm to improve the classification accuracy of heart disease data set. We used genetic search as a goodness measure to prune redundant and irrelevant attributes, and to rank the attributes which contribute more towards classification. Least ranked attributes are removed, and classification algorithm is built based on evaluated attributes. This classifier is trained to classify heart disease data set as either healthy or sick. Our proposed algorithm consists of two parts.
1) First part deals with evaluating attributes using genetic search
2) Part two deals with building classifier and measuring accuracy of the classifier
Proposed algorithm is shown in fig 5

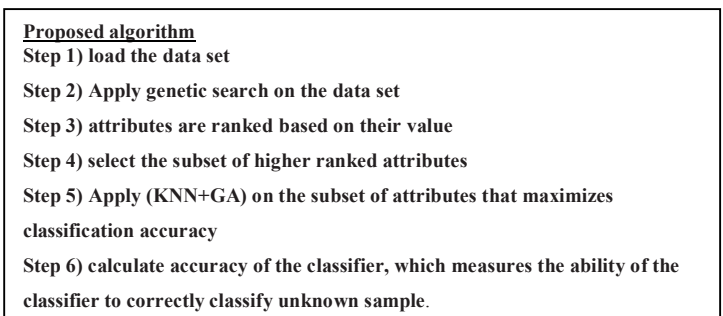

**Proposed algorithm**
**Step 1) load the data set**

**Step 2) Apply genetic search on the data set**

**Step 3) attributes are ranked based on their value**

**Step 4) select the subset of higher ranked attributes**

**Step 5) Apply (KNN+GA) on the subset of attributes that maximizes classification accuracy**

**Step 6) calculate accuracy of the classifier, which measures the ability of the classifier to correctly classify unknown sample**.

**Fig 5 our proposed algorithm**

**Step 1 to 4** comes under part 1 which deals with attributes and their ranking.
**Step 5** is used to build the classifier and **step 6** records the accuracy of the classifier.

Accuracy of the classifier is computed as
Accuracy =    no. Of samples correctly classified in test data
              ----------------------------------------------------------
              Total no.of samples in the test data

## 4. Results and discussion

The performance of our proposed approach has been tested with 6 medical data sets and 1 non medical data set. Out of 7 data sets, 6 data sets were chosen from UCI Repository [12] and heart disease A.P was taken from various corporate hospitals in Andhra Pradesh, and attributes are selected based on opinion from expert doctor's advice. Information about these attributes is listed in table 1.As a way to validate the proposed method, we have tested with emphasis on heart disease on A.P besides other machine learning data sets taken from UCI repository. In table 1, column 1 represents name of the attribute, column 2 no. of instances, and column 3 represents no. Of attributes in each data set .The comparison of our proposed algorithm with 4 algorithms is listed in table 2.Accuracy of 5 data sets is improved by our approach. Accuracy of various data sets using full training sets is represented in table 3.We carried out experiments by assigning various values K.As the k value goes on increasing accuracy of data sets is decreasing. Accuracy of various data sets using the cross validation with genetic algorithm is shown in table 4.accuracy of heart disease data decreases 32% using cross validation. We used 5 fold cross validation. Accuracy of the data sets with and without GA is shown in table 5 and 6.Accuracy of the heart disease is increased by 5% using and GA using full training data set and 15%improvement in accuracy for cross validation against KNN without GA. Accuracy of hypothyroid has been improved 2.94% against classification accuracy of NN+PCA [13].Attributes of heart disease and their corresponding data type is shown in table 7.Our method (KNN+GA) was not successful for breast cancer and primary tumor .This may due to our proposed approach could not account for irrelevant and redundant attributes present in above mentioned data sets. Cross over rate for GA   should be high and 60% is preferable ,so we set the value at 60%.Mutation rate should be low and we set mutation value at 0.033.Population size should be good to improve the performance of GA,so population size is fixed at 20.

The results acquired by KNN and GA reveals that by integrating GA with KNN will improve the classification accuracy for many data sets and especially heart disease for A.P.

**Table 1   Description of various data sets**

| Data set | Instances | Attributes |
|---|---|---|
| Weather | 14 | 5 |
| Pima | 768 | 9 |
| hypothyroid | 3770 | 30 |
| breast cancer | 286 | 10 |
| liver disorder | 345 | 7 |
| primary tumor | 339 | 18 |
| heart stalog | 270 | 14 |
| lymph | 148 | 19 |

**Table 2 Accuracy comparison with various algorithms**

| Data set name | KNN+GA(our approach) | NN+PCA | NN+ $\chi^2$ | GA+ANN |
|---|---|---|---|---|
| Weather data | 100 | 100 | 100 | 100 |
| Breast cancer | 94 | 97.9 | 97.6 | 95.45 |
| Heart stalog | 100 | 98.14 | 97.7 | 99.6 |
| Lympography | 100 | 99.3 | 100 | 99.3 |
| Hypothyroid | 100 | 97.06 | 97.64 | 97.37 |
| Primary tumor | 75.8 | 80 | 83.1 | 82.1 |
| Heart disease A.P | 100 | 100 | 100 | 100 |

**Table 3 Accuracy of various data sets using full training set with GA**

| Data set name | K=1 | K=3 | K=5 |
|---|---|---|---|
| Weather data | 100 | 100 | 100 |
| Breast cancer | 94.05 | 94.05 | 94.05 |
| Heart stalog | 100 | 90.7 | 87.03 |
| Lympography | 100 | 100 | 100 |
| Hypothyroid | 100 | 96.18 | 95.75 |
| Primary tumor | 75.8 | 66.3 | 60.7 |
| Heart disease A.P | 100 | 95 | 95 |

**Table 4 Accuracy of various data sets using cross validation with GA**

| Data set name | K=1 | K=3 | K=5 |
|---|---|---|---|
| Weather data | 78.57 | 78.57 | 78.57 |
| Breast cancer | 73.07 | 71.32 | 74.47 |
| Heart stalog | 80 | 78.14 | 81.4 |
| Lympography | 85.13 | 85.13 | 87.1 |
| Hypothyroid | 91.7 | 93.47 | 94.0 |
| Primary tumor | 40.7 | 45.13 | 47.4 |
| Heart disease A.P | 67.5 | 65 | 62.5 |

**Table 5 Accuracy of various data sets using full training set without GA**

| Data set name | K=1 | K=3 | K=5 |
|---|---|---|---|
| Weather data | 85.71 | 85.71 | 85.71 |
| Breast cancer | 90 | 90 | 82.5 |
| Heart stalog | 100 | 90.74 | 83.3 |
| Lympography | 99.32 | 99.32 | 84.4 |
| Hypothyroid | 100 | 95.62 | 94.69 |
| Primary tumor | 75 | 65.48 | 61.35 |
| Heart disease A.P | 95 | 75 | 83.3 |

**Table 6 Accuracy of various data sets using cross validation without GA**

| Data set name | K=1 | K=3 | K=5 |
|---|---|---|---|
| Weather data | 57.4 | 64.28 | 57.14 |
| Breast cancer | 73.07 | 71.32 | 72.7 |
| Heart stalog | 72.96 | 77.47 | 78.8 |
| Lympography | 81.08 | 82 | 84.4 |
| Hypothyroid | 91.46 | 93.29 | 93.45 |
| Primary tumor | 39.82 | 43.06 | 46.01 |
| Heart disease A.P | 52.5 | 50 | 57.5 |

**Table 7 Attributes of heart disease**

| Sl.no | Attribute | Data Type |
|---|---|---|
| 1 | Age | Numeric |
| 2 | Gender | Nominal |
| 3 | Diabetic | Nominal |
| 4 | BP Systolic | Numeric |
| 5 | BP Dialic | Numeric |
| 6 | Height | Numeric |
| 7 | Weight | Numeric |
| 8 | BMI | Numeric |
| 9 | Hypertension | Nominal |
| 10 | Rural | Nominal |
| 11 | Urban | Nominal |
| 12 | Disease | Nominal |

**KNN PARAMETERS**
1) K=1,2,3---N
2) Cross validate=True
3) Debug =True
4) Ditance weighting=weight by 1 distance
5) Mean squared=True
6) No normalization=False

**Fig 6 KNN Parameters**

**GENETIC SEARCH PARAMETERS**
1) Cross over probability=0.6
2) Mutation probability=0.033
3) Maximum generations=20
4) Population size=20
5) Report frequency=20
6) Seed =1

**Fig 7 Genetic search parameters**

**Table 8 Accuracy comparison with and without GA**

| Data set name | Accuracy without GA(knn only) | Accuracy with GA(knn+GA) |
|---|---|---|
| Weather data set | 85.71 | 100 |
| Breast cancer | 90 | 94.35 |
| Heart stalog | 100 | 100 |
| Lympography | 99.32 | 100 |
| Hypothyroid | 100 | 100 |
| Primary tumor | 75 | 75.8 |
| Heart disease A.P | 95 | 100 |
| Average | 92.14 | 95.73 |

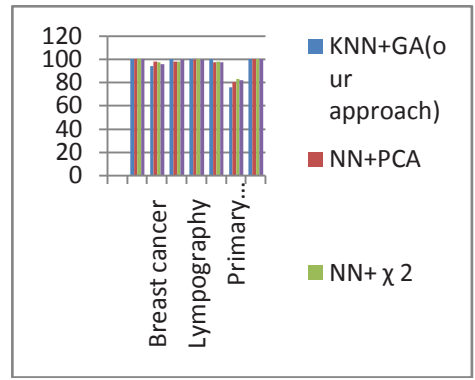

**Fig 8 Accuracy comparision of various data sets**

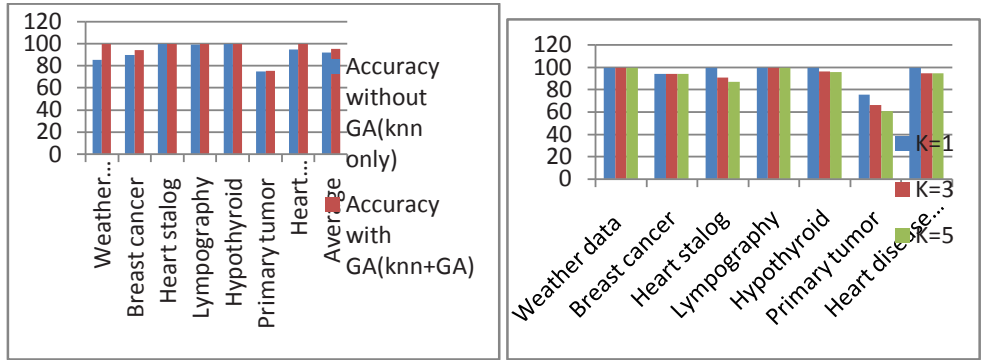

**Fig 9 Accuarcy comparision with and without GA**   **Fig 10 Accuracies of various data sets for different K values**

Parameters of KNN and genetic search are described figure 6 and figure 7. Figure 8 to figure 10 shows accuracy comparision of various data set.Table 8 shows accuracy of various data sets with and without GA.Average accuracy of our approach is higher than KNN approach with out GA.Accuracy of heart disease A.P is improved 5% over classification algorithm with out GA.Accuracy of weather data set is 14.29% by our approach. From the results it is also observed that integrating GA with KNN out performs the other methods with greater accuracy.

## 5. Conclusion

In this paper we have presented a novel approach for classifying heart disease. As a way to validate the proposed method, we have tested with emphasis on heart disease on A.P besides other machine learning data sets taken from UCI repository. Experimental results carried out on 7 data sets show that our approach is a competitive method for classification. This prediction model helps the doctors in efficient heart disease diagnosis process with fewer attributes. Heart disease is the most common contributor of mortality in India and in Andhra Pradesh. Identification of major risk factors and developing decision support system, and effective control measures and health education programs will decline in the heart disease mortality.